\documentclass[aps,prl,twocolumn,showpacs,superscriptaddress,showkeys,bm,amsmath,amssymb]{revtex4-1}
\usepackage{graphicx}\usepackage{xcolor}\usepackage{dcolumn}
\usepackage[normalem]{ulem}
\begin{document}
\title{Extended Axion Dark Matter Search Using the CAPP18T Haloscope}
\newcommand{\SNU}{\affiliation{Department of Physics and Astronomy, Seoul National University, Seoul 08826, Korea}}
\newcommand{\KAIST}{\affiliation{Department of Physics, Korea Advanced Institute of Science and Technology, Daejeon, 34141, Korea}}
\newcommand{\IBS}{\affiliation{Center for Axion and Precision Physics Research, Institute for Basic Science, Daejeon, 34051, Korea}} 
\author{Byeongsu Yang}\SNU
\author{Hojin Yoon}\KAIST
\author{Moohyun Ahn}\SNU
\author{Youngjae Lee}\KAIST
\author{Jonghee Yoo}\SNU\KAIST\IBS
\date{\today}
\begin{abstract}
We report an extended search for the axion dark matter using the CAPP18T haloscope. The CAPP18T experiment adopts innovative technologies of a high-temperature superconducting magnet and a Josephson parametric converter. The CAPP18T detector was reconstructed after an unexpected incident of the high-temperature superconducting magnet quenching. The system reconstruction includes rebuilding the magnet, improving the impedance matching in the microwave chain, and mechanically readjusting the tuning rod to the cavity for improved thermal contact. The total system noise temperature is $\sim$0.6\,K. The coupling between the cavity and the strong antenna is maintained at $\beta \simeq 2$ to enhance the axion search scanning speed. The scan frequency range is from 4.8077 to 4.8181 GHz. No significant indication of the axion dark matter signature is observed. The results set the best upper bound of the axion-photon-photon coupling ($g_{a\gamma\gamma}$) in the mass ranges of 19.883 to 19.926\,$\mu$eV at $\sim$0.7$\times|g_{a\gamma\gamma}^{\text{KSVZ}}|$ or $\sim$1.9$\times|g_{a\gamma\gamma}^{\text{DFSZ}}|$ with 90\,\% confidence level. The results demonstrate that a reliable search of the high-mass dark matter axions can be achieved beyond the benchmark models using the technology adopted in CAPP18T.
\end{abstract}
\maketitle

\par Axions are hypothetical particles introduced to solve the {strong {\it CP}} problem. The axions may have been {nonthermally} produced in the early Universe\,\cite{Planck, Abbott:1982af, DINE1983137, Preskill:1982cy, PhysRevLett.50.925, Abbott:1982af, DINE1983137,Sikivie:2006ni,PhysRevD.78.083507}. Cosmological investigation suggests that the axion mass is on the scale of $\mu$eV or higher\,\cite{Bonati2016, PhysRevD.92.034507, Borsanyi2016, PhysRevLett.118.071802, PhysRevD.96.095001, PETRECZKY2016498}. These axions extremely weakly interact with normal matter through anomalous coupling. Therefore, the axions are regarded as excellent candidates for dark matter. Two benchmark axion models are outstanding; the Kim-Shifman-Vainshtein-Zakharov (KSVZ) model\,\cite{Kim:1979if, Shifman:1979if} and the Dine-Fischler-Srednicki-Zhitnitsky (DFSZ) model\,\cite{Dine:1981rt,Zhitnitsky:1980tq}. While a vast range of {\it QCD axion models} {has} remained unexplored\,\cite{DiLuzio:2020wdo}, a part of the mass to coupling parameter space of the classical {\it invisible axion models} is ruled out, {in particular, the} $\mu$eV to meV range {remains} open to be explored. There are {now} tremendous efforts to experimentally probe the axion dark matter\,\cite{Bartram:2020ysy,Braine:2019fqb,Du:2018uak,Asztalos:2009yp,Asztalos:2003px,Hagmann1998a, Brubaker:2016ktl,Brubaker:2017rna,Backes2021a, Alesini2019a,Alesini2021a, Lee:2020cfj,Jeong:2020cwz, Kwon2021a}. Haloscopes are one of the most competitive detector {technologies} for the axion dark matter search\,\cite{Sikivie:1983ip}. A standard axion haloscope is equipped with a resonant microwave cavity in a strong solenoid magnet. The virtual photons of the magnetic field in the cavity may couple to the dark matter axions and convert them to microwave photons which can be detected using {radio frequency (rf)} technology.

\par CAPP18T is an axion haloscope that adopted innovative technologies of an 18\,T high-temperature superconducting ({high-temperature superconducting}) solenoid magnet and a quantum-limited Josephson {parametric converter} (JPC). Details of the CAPP18T detector can be found in Ref.\,\cite{capp18t:detector}. In the first phase axion dark matter search operation in 2020 ({run 1}), the best axion dark matter bound was set in the mass range of 19.764 {--} 19.890\,$\mu$eV\,\cite{Lee:2022mnc}. However, an unexpected magnet quenching incident occurred during the scheduled {shutdown} of the detector on {24 December} 2020\,\cite{capp18t:quench}. The magnet and the cavity were damaged during the incident, while other detector parts were intact. The haloscope was completely dismantled and reassembled; the magnet vendor (SuNAM Co. Ltd., Korea) repaired the 18\,T magnet, and the cavity was replaced with a new one. The thermal contact between the JPC and the mixing chamber of the dilution {refrigerator was} improved using a newly designed detector frame. The impedance of the cryogenic {rf} chain was thoroughly reviewed and renewed. This Letter reports the axion dark matter search results with this reconstructed CAPP18T detector.\\

\par The axion interaction in the CAPP18T is expected to occur in a {1-lL-volume} cylindrical copper cavity. A copper rod in the cavity adjusts the cavity resonant frequency $\nu_C$. A strong coupling antenna at the cavity top receives the {rf} signals, where the coupling ($\beta$) is controlled by altering the insertion depth of the antenna. The sensitivity of the axion search is determined by the signal-to-noise ratio (SNR = ${P_{a}}\sqrt{{t}/{\Delta \nu}}/({k_{B}{T_\text{syst}})}$), where $P_a$ is the axion signal power, $t$ is the scan time, $\Delta \nu$ is the signal bandwidth, $k_B$ is the Boltzmann constant, and ${T_\text{syst}}$ is the system noise temperature. The axion signal power is given by
\begin{eqnarray}
P_a = g^2_{a\gamma\gamma}\Big(\dfrac{\rho_a}{m_a}\Big)B^2 V C \dfrac{\beta}{(1+\beta)} Q_L,
\label{eq:Pax}
\end{eqnarray}
\noindent where $g_{a\gamma\gamma}$ is the axion-photon-photon coupling, $\rho_a$ is the local dark matter density, $m_a$ is the axion mass, $B$ is the external magnetic field, $V$ is the cavity volume, $Q_L$ is the loaded quality factor of the cavity, and $C$ is the form factor{, which is unitless and of order one}. After the full recovery and improvement of the CAPP18T detector, the dark matter search experiment was carried out from June to August 2021 ({run 2}, this report). The experiment is targeted to seek axions with the KSVZ coupling in the frequency range of 4.8077 {-- 4.8181}\,GHz. Data samples of the corresponding range are combined into 78 groups by their target frequencies. The difference between each adjacent target frequency is maintained between 100 {and} 200\,kHz. Each group is 12 {h} of exposure data, divided into {six} subgroups of 2 {h}. Each subgroup consists of 10 spectra of 12 {min}. A new target $\nu_C$ is set every 12 {h}.\\

\par Adjustment of the {rf} system is made before the automatic data acquisition. $\nu_C$ and $\beta$ are manually adjusted using a {vector network analyzer}. $P_\text{off}$, the noise power with the JPC off, is then measured with a digitizer for 100 {sec}. These data are used for the noise temperature calibration. The JPC resonant frequency $\nu_{J}$ is set close to $\nu_C$, and the JPC gain $G_{J}(\nu)$ at $\nu = \nu_{J}$ is adjusted to $\sim$27\,dB. An automated DAQ program then acquires {2-h-long} subgroup data. The DAQ program measures $\nu_C$ and $\beta$, and performs an automatic {readjustment} if the differences between the measured and targeted parameters exceed the {retuning} tolerance criteria of $|\Delta\nu_C| < 10$\,kHz and $|\Delta\beta| < 0.05$. The typical values (root-mean-squared) of $Q_L$ and $\beta$ are about 14\,300 (18.2\,\%) and 2.0 (1.4\,\%), respectively. The JPC is turned on during the noise power measurements ($P_\text{on}$). A calibration signal is injected at $\nu_C$ for a few seconds. The calibration signal frequency is then shifted away at $\nu_C$+500\,kHz, and $P_\text{on}$ is acquired for 12 {min}. The above data acquisition procedure is repeated {10} times. Additional $G_{J}$ and $P_\text{off}$ measurements are then carried out, which completes a subgroup data-taking process. If drift in $\nu_{J}$ occurs, $\nu_{J}$ and $G_{J}$ are readjusted in the subsequent subgroup measurement. The DAQ efficiency is improved by about 15\,\% compared to the previous operation. This improvement is achieved by automating the DAQ processes. The scan speed is given by $d\nu/dt \propto \beta^2/(1+\beta)^2 Q_L = \beta^2/(1+\beta)^3 Q_0$\,\cite{capp18t:detector}, where $Q_0$ is the unloaded quality factor of the cavity. {Note that we set $\beta \simeq 1$ in {run 1} and $\beta \simeq 2$ in {run 2}.} Ideally, the scan speed is maximized at $\beta = 2$, {expecting $\sim$20\,\% of improved} scan speed than the case of $\beta = 1$. {However, we achieved only} about 10\,\% of the overall scan speed improvement in practice, mainly due to the correlation between the $Q_L$ and the $\beta$ value. $C$ is {estimated from simulation to be between 0.58 and 0.59} in the scanned frequency region. \\

\par The system noise temperature ${T_\text{syst}}$ is measured using the $Y$-factor method and the SNR {improvement} (SNRI) method\,\cite{Du:2018uak}. The relation between ${T_\text{syst}}$ and SNRI is given by
\begin{equation}
{T_\text{syst}} =  \dfrac{{T_\text{syst}^\text{off}}}{\mathrm{SNRI}} = {T_\text{syst}^\text{off}}   \frac{G_\text{off}}{G_\text{on}}\frac{P_\text{on}}{P_\text{off}},
\end{equation}
\noindent where $G_{\text{on} \text{(off)}}$ is the transfer function when the JPC is on (off), ${T_\text{syst}^\text{off}}$ is the system noise temperature when the JPC is off, and $\mathrm{SNRI}$ is defined as $(G_\text{on}P_\text{off})/(G_\text{off}P_\text{on})$. The $Y$-factor method consists of pairs of noise spectra measured from the cold and hot load. For the cold load power, $P_\text{off}$ is used. A hot-load measurement is performed every 22 {h} before the liquid helium recharges in the cryostat. The hot load, an {rf} terminator, is placed in the 4\,K stage. The receiver's input is connected to the hot load by inverting the receiver switch for the {hot-load} measurement. As the temperature of the whole system rises due to the heat generated by the receiver switch, a {30 min} interval is given until the temperature cools down. The thermal photons from the hot load are then measured for 100 {sec} every 10 {min} until the noise values are stabilized. The {hot-load} temperature $T_\text{hot}$ varied between 4.9 {and} 5.2 K. The measured ${T_\text{syst}^\text{off}}$ is 5.32$\pm$0.28\,K.\\

\begin{figure}[t!]
\centering
\includegraphics[width=1\linewidth]{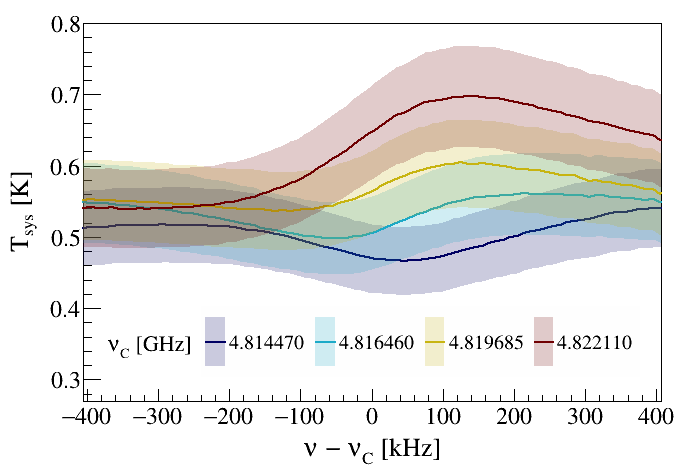}
\caption{Typical noise temperature spectra as a function of $\nu - \nu_C$. The spectra show dip or asymmetric peak shapes.}
\label{fig:pon}
\end{figure}

\par
\begin{figure}[t!]
\centering
\includegraphics[width=1\linewidth]{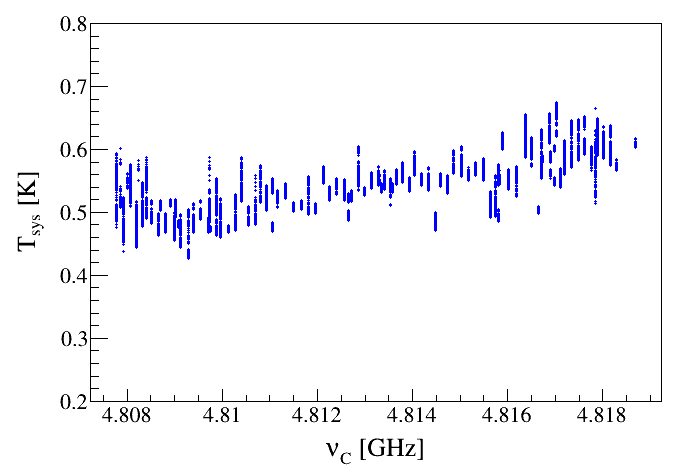}
\caption{Measured ${T_\text{syst}}$ for each $\nu_C$. {Each data point is derived from individual spectra measurements.}}
\label{fig:Tsys}
\end{figure}

 The SNRI is measured for each $P_\text{on}$. The ratio of the total gain with the JPC on and off, $G_\text{on}/G_\text{off}$, is obtained from $G_\text{on}/G_\text{off} = G_{J}/\epsilon$, where $\epsilon$ is the attenuation between the JPC and the {high electron mobility transistor} (HEMT, LNF-LNC2-6A). The SNRI varies between 9 {and} 11\,dB depending on $\nu - \nu_C$, where $\nu$ is the measured frequency. Figure \ref{fig:pon} shows a set of typical noise temperature spectra as a function of $\nu - \nu_C$. One sample shows a dip around the cavity resonant frequency, while others show asymmetric peak shapes. The origin of these shapes is not clearly understood, and we suspect impedance mismatching or variation in the {rf} system. Figure \ref{fig:Tsys} shows ${T_\text{syst}}$ as a function of $\nu_C$. {Each point in Figure \ref{fig:Tsys} corresponds to the ${T_\text{syst}} (\nu)$ value at $\nu = \nu_C$ of their corresponding measured spectrum.} The parameters relevant to ${T_\text{syst}}$ of {run 2} are compared with the {run 1} values in Table~\ref{tab:Tsyspar}. The mixing chamber temperature is $\sim$60\,mK, the cavity temperature is maintained at $\sim$210\,mK, and the physical JPC temperature throughout the experiment is at $\sim$70\,mK. After the {run 1} incident, the HEMT was replaced. The gain of the HEMT in {run 2} is 33.3\,dB, reduced by $\sim$3.4dB compared to that of {run 1}; this reduced the typical $P_\text{off}$ by half. The thermal contact of the detector components, especially the JPC to the mixing chamber, was carefully inspected to lower ${T_\text{syst}}$. The typical ${T_\text{syst}}$ of this run is $\sim$535\,mK, which is lower than that of {run 1} ($\sim$840\,mK).\\

\begin{figure*}[t!]
\includegraphics[width=1\linewidth]{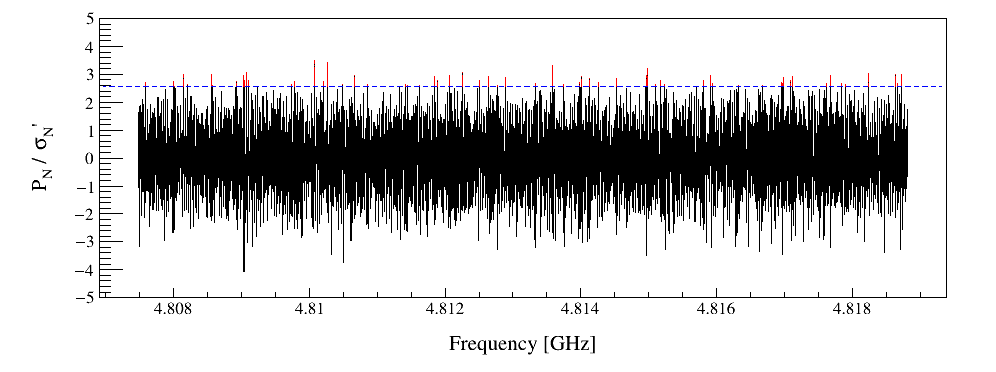}
\caption{Scaled grand power spectrum {($P_N/\sigma_N'$)}. The threshold at 2.550 is shown {with} the blue dotted line. The 58 signal peaks exceeding the threshold are marked in red.}
\label{fig:pn}
\end{figure*}

\begin{table}[b!]
\caption{\label{tab:Tsyspar} Comparison of typical detector parameters of {run 1} and {run 2}. The decrease of $P_\text{on}$ and the increase of SNRI are the main sources of the reduced ${T_\text{syst}}$ and the JPC noise temperature ($T_{J}$) in {run 2}.}
\begin{ruledtabular}
\begin{tabular}{l c c}
Parameter & {Run 1} (2020) & {Run 2} (2021)  \\
\hline
MC T [mK]                     & 60                & 60  \\
Cav T [mK]                    & 110               & 210 \\
${T_\text{syst}}^\text{off}$ [K] & 4.7$\pm$0.2       & 5.3$\pm$0.3 \\
$P_\text{off}$ [pW]           & 57$\pm$2          & 26$\pm$1  \\ 
$P_\text{on}$ [nW]            & 4.7$\pm$0.9       & 1.3$\pm$0.2  \\ 
JPC Gain [dB]                 & 27                & 27   \\
JPC noise $T_{J}$ [mK]        & 503 (2.2 photons) & 292 (1.3 photons)\\
HEMT Gain [dB]                & 36.7              & 33.3      \\
SNRI [dB]                     & 7.5$\pm$0.3       & 9.5$\pm$0.4 \\
${T_\text{syst}}$ [mK]           & 840$\pm$70        & 540$\pm$50                  
\end{tabular}
\end{ruledtabular}
\end{table}

\par The analysis procedure is analogous to Refs.\,\cite{Lee:2022mnc,capp18t:detector}. The temperatures of the detector components are kept stable, and no abrupt change in the SNRI is observed. The initial data selection criteria remove 0.8\,\% of the data {: nontypical} SNRI, $Q_L$ fluctuation over 1\,000, and $\nu_C$ drift over 10\,kHz. The spectrum baseline is subtracted and divided by the Savitzky-Golay (SG) filter baseline of degrees 3 and 5\,001 data points. The spectrum is then scaled with ${T_\text{syst}}$ and {divided by} $P_a$ to obtain the {rescaled} power. All spectra are combined into a single spectrum using the weighted average method{, where the weight is the inverse of the corresponding variance for each frequency bin}. The combined spectrum is then rebinned by {integrating it into a 5 kHz bin for presentation{, which is the grand spectrum, $P_N$}. The actual likelihood calculation was performed by running 5 kHz windows in 10 Hz resolution, in which each likelihood value was obtained by shifting the window {10 Hz} from the previous one. The choice of the 5 kHz window is set to maximize {the} SNR for the boosted Maxwell-Boltzmann (MB) model\,\cite{PhysRevD.42.3572}.}
{A linear correlation between the standard deviation of $P_N/\sigma_N$ and the number of the combined spectra was observed, where $\sigma_N$ is the weighted standard deviation of $P_N$. As more spectra were added, the standard deviation of $P_N/\sigma_N$ also linearly increased. The correlation is caused by the SG filter; the filter may not have completely removed the baseline structure, and the residual shape remains. In {run 1}, the target LO frequency varied from {subrun to subrun}, where each {subrun} duration was about tens of minutes. Conversely, in {run 2}, each target LO frequency is fixed for 12 {h}. This long duration at a fixed frequency caused an increase in the standard deviation of $P_N/\sigma_N$ from 1 to 1.458. Considering the linear correlation, $\sigma_N$ was scaled to $\sigma_N’ = \zeta \sigma_N$ where $\zeta$ is a scale factor to normalize $P_N/\sigma_N’$ to 1. Figure\,\ref{fig:pn} shows the scaled grand power spectrum $P_N/\sigma_N’$ on which the axion signal search is performed.}
The correlation was confirmed in the MC study, which affected the performance of the SG filter, resulting in about 30\,\% SNR degradation. Consequently, the SNR reduction by the SG filter is estimated to be 0.609. {{Because of} the correlation, the signal threshold (the rescan threshold) is lowered, which means the number of rescan candidates is increased compared to the higher threshold case. Therefore, the detector operation time for the rescan is increased. As a result, the axion search sensitivity is not significantly affected by the degradation.} This is because while $\sigma_N$ is increased {to $\sigma_N'$}, $P_N$ derived from the measured values remains unaffected. The target SNR is lowered to 3.832, which gives the signal threshold of 2.550 with a 90\,\% confidence level. {The target SNR refers to the ratio of the target intensity of the signal to a given noise fluctuation if the axions are present. The target threshold refers to the minimum SNR value in which the presence of an axion signal can be claimed.} Note that in {run 1}, the target axion SNR (threshold) was at 5.0 (3.718)\,\cite{capp18t:detector}.  While $P_N/\sigma_N$ is scaled down {to $P_N/\sigma_N'$}  by 1/{$\zeta$} = 0.686, the target SNR is scaled down by 3.832/5.0 = 0.766, which eventually reduced the overall sensitivity by 6\,\%. A total of 58 candidates exceeded the target threshold. These primary candidates are tested by the likelihood method. {The signal-like probability $p_A$ is calculated based on the probability distributions with the {boosted} MB model\,\cite{PhysRevD.42.3572}. Similarly, the {noiselike} probability $p_N$ is obtained based on a Gaussian white noise distribution. Therefore, the value of $p_A+p_N$ is not necessarily unity as they are estimated using independent distributions.} The signals {with $p_A$} over 1\,\% are the final candidates for the rescanning experiment. Table~\ref{tab:candidates} shows the list of rescanning frequencies, excess significance $\sigma$, $p_A$, {and $p_N$}. The SNR is monitored during the rescanning experiment; the rescan is stopped if the candidate disqualifies the conditions of an axion signal. All 20 rescan candidates failed the test, convincing {us} that they do not originate from the axion dark matter.

\begin{table}[b!]
\caption{\label{tab:candidates} List of likelihood test results of candidates with $p_A > 1\,\%$. Rescan operations are performed on these 20 candidates.}
\begin{ruledtabular}
\begin{tabular}{c l c c c c}
No. & $\nu_a$ [GHz] & excess [$\sigma$] & $p_A$ & $p_N$ \\
\hline
 1 & 4.808\,924 & 2.772 & 0.0213 & 0.4453 \\
 2 & 4.809\,032 & 2.965 & 0.1276 & 0.1469 \\
 3 & 4.809\,077 & 3.055 & 0.3386 & 0.0472 \\
 4 & 4.809\,102 & 2.775 & 0.1287 & 0.1449 \\
 5 & 4.809\,741 & 2.663 & 0.2590 & 0.0673 \\
 6 & 4.811\,623 & 2.584 & 0.3101 & 0.0544 \\
 7 & 4.812\,505 & 2.790 & 0.3403 & 0.0467 \\
 8 & 4.812\,642 & 2.918 & 0.1775 & 0.1083 \\
 9 & 4.813\,338 & 2.652 & 0.3205 & 0.0517 \\
10 & 4.813\,587 & 3.348 & 0.1182 & 0.1646 \\
11 & 4.814\,522 & 2.852 & 0.2665 & 0.0654 \\
12 & 4.815\,236 & 2.654 & 0.3424 & 0.0462 \\
13 & 4.815\,809 & 2.865 & 0.1675 & 0.1156 \\
14 & 4.816\,666 & 2.636 & 0.1176 & 0.1659 \\
15 & 4.816\,960 & 2.753 & 0.2509 & 0.0694 \\
16 & 4.816\,984 & 2.903 & 0.0335 & 0.3690 \\
17 & 4.817\,112 & 2.901 & 0.1147 & 0.1713 \\
18 & 4.817\,837 & 2.745 & 0.2634 & 0.0662 \\
19 & 4.818\,634 & 3.048 & 0.2851 & 0.0607 \\
20 & 4.818\,715 & 3.062 & 0.1373 & 0.1376 \\
\end{tabular}
\end{ruledtabular}
\end{table}

\par The bounds of the axion dark matter are set in two ways; the {frequentist} approach, which follows Ref.\,\cite{Brubaker:2017rna}, and the Bayesian approach, as used in Refs.\,\cite{Palken:2020wgs,Bartram:2020ysy,Cervantes:2022epl,feldman1998unified,capp18t:detector}. In the Bayesian method, the posterior probability for the expected signal power $\mu_a$ given the measured power $P_N$ is
\begin{equation}
    \text{P}(\mu_a|P_N) \sim \text{exp}\left [ -\frac{1}{2}\left ( \frac{P_N - \mu_a}{\sigma_N}\right ) ^2\right ] \Theta (\mu_a), 
\end{equation}
\noindent where $\Theta(\mu_a)$ is a step function~\,\cite{Cervantes:2022epl,feldman1998unified}. From Bayes's theorem, $\text{P}(\mu_a|P_N) = \text{P}(P_N|\mu_a)\text{P}(\mu_a)/\text{P}(P_N)$, where $\text{P}(P_N|\mu_a)$ is a Gaussian, the probability of measuring $P_N$ for a signal at $\mu_a$, and $\text{P}(P_N)$ is a normalization factor. The prior assumption is that only a physical $g_{a\gamma\gamma}$ value should be regarded, or that $\mu_a$ should not be a negative value with uniform probability, which gives $\text{P}(\mu_a) \sim \Theta(\mu_a)$. This step function results in a truncated Gaussian distribution function. {The cumulative distribution function for $\text{P}(\mu_a|P_N)$ is then used to determine the sensitivity limit at the 90\,\% {credibility level}.} The systematic uncertainties related to the axion signal power from the cavity are listed in Table~\ref{tab:SysUncert}. The total systematic uncertainty is 11.0\,\%.

\begin{figure*}[t]
\includegraphics[width=1\linewidth]{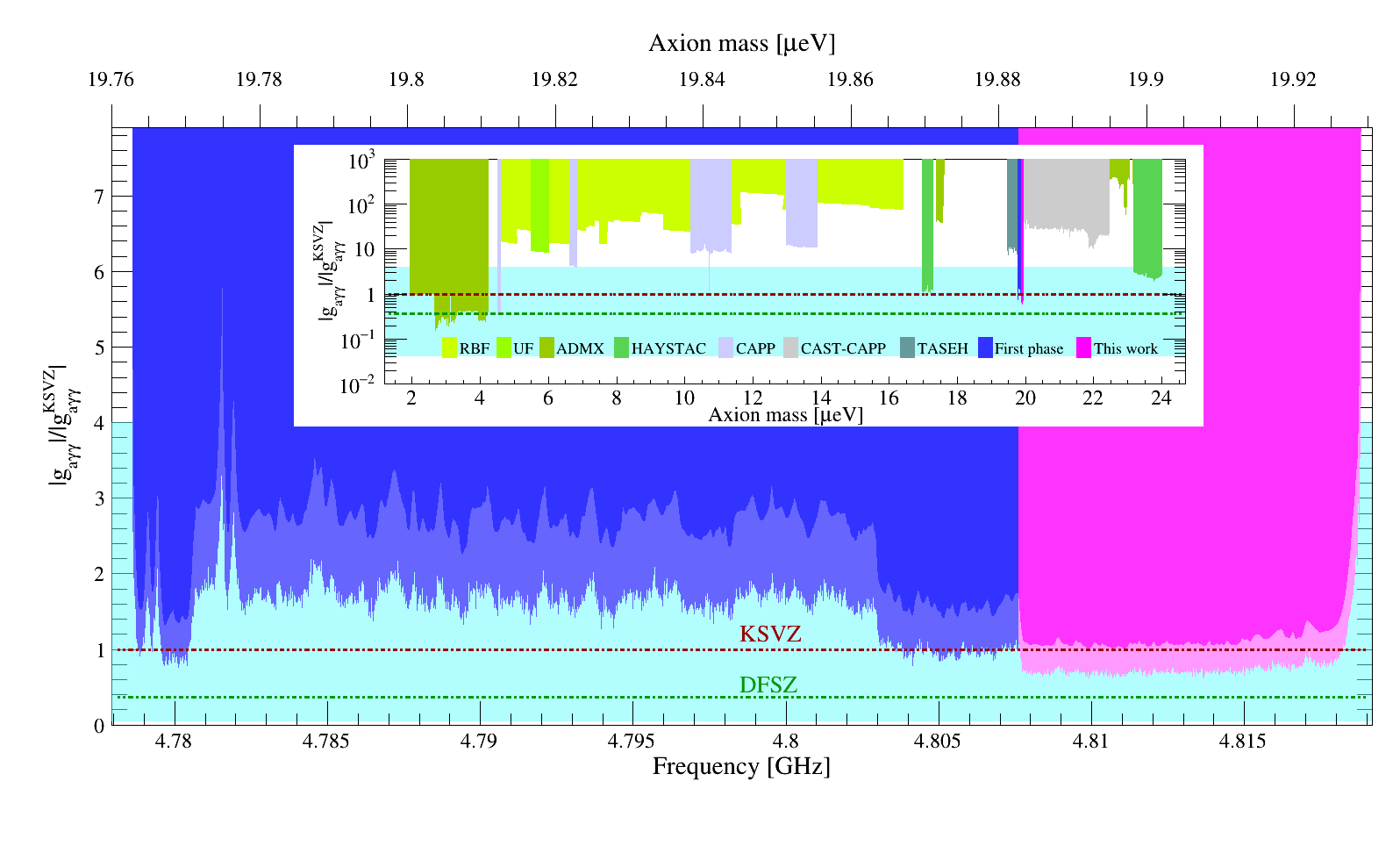}
\caption{Upper bound of axion-photon-photon coupling of the CAPP18T axion dark matter search. The light magenta region shows the exclusion limit at 90\,\% {Credibility Level} from this work in the Bayesian method. The magenta region shows the limit in the {frequentist} method. Light blue and blue show the Bayesian and {frequentist} results from the previous CAPP18T work, respectively\,\cite{Lee:2022mnc}. The theoretical benchmark models (KSVZ and DFSZ) are shown as {dotted brown} and {dotted green} lines with the uncertainty band (light cyan)\,\cite{Cheng1995a}. The inset shows the upper bounds of our works (magenta and blue) together with other haloscope results: ADMX (olive, Refs.\,\cite{Hagmann1998a,Asztalos:2001tf,Asztalos2002a, Asztalos:2003px, Asztalos:2009yp,Sloan2016a, Boutan2018a,ADMX:2021nhd}, C.L.\,=\,95\,\%), HAYSTAC (green, Refs.\,\cite{Brubaker:2016ktl, Backes2021a, PhysRevD.97.092001}, C.L.\,=\,90\,\%), {CAPP (light purple, Refs.\,\cite{Lee:2020cfj, Jeong:2020cwz, Kwon2021a,kim2023near,andrew2023axion}, C.L.\,=\,90\,\%), CAST-CAPP (gray, Ref. \cite{adair2022search}), and TASEH (dark green, Ref. \cite{chang2022first}).} RBF (lime, Refs.\,\cite{DePanfilis1987a, Wuensch1989a}, C.L.\,=\,95\,\%) and UF (light green, Ref.\,\cite{Hagmann1990a}, C.L.\,=\,95\,\%) limits shown here are rescaled based on $\rho_a$\,=\,0.45 GeV/cm$^3$.}
\label{fig:limit}
\end{figure*}

\par Figure\,\ref{fig:limit} shows the exclusion limit for $g_{a\gamma\gamma}$ over the scanned mass range set by the Bayesian method, along with the {frequentist} method. The sensitivity limit is shown every 20\,kHz for {the} Bayesian, while every 10\,Hz for {the frequentist method}. The following procedures are applied to smooth the Bayesian sensitivity curve. For every grand spectrum bin ($P_N$,$\sigma_N$), 100 Monte Carlo random samples are generated following the truncated Gaussian distribution. Every 2\,000 consecutive bins are grouped into 20\,kHz width, and a 90\,\% {C.L.} for each 200\,000 sample group is then determined. The Bayesian method gives over a 35\,\% better result than the {frequentist} method. Assuming the {boosted} {MB} distribution of dark matter{\,\cite{PhysRevD.42.3572}} and the local dark matter density of 0.45 GeV/cc (100\,\% axions) {from the standard halo model}, {axions with coupling over $\sim$0.7$\times|g_{a\gamma\gamma}^{\text{KSVZ}}|$ or  $\sim$1.9$\times|g_{a\gamma\gamma}^{\text{DFSZ}}|$ were excluded} in the mass range of 19.883 -- 19.926\,$\mu$eV (90\,\% {credibility level}). \\

\begin{table}[b!]
\caption{Systematic uncertainties associated with the axion signal power from the cavity.}
\begin{tabular}{l c r}
\hline 
\hline 
~~{Source}       &{~~~~~~~~~~~~}&{Fractional uncertainty on $P_a$~~}\\
\hline
~~$B^2 V$        &  &1.4\%~~~~~~~~~~~~~~~~~ \\
~~$Q_L$          &  &0.6\%~~~~~~~~~~~~~~~~~ \\
~~$\beta$        &  &0.2\%~~~~~~~~~~~~~~~~~ \\
~~$C$            &  &3.9\%~~~~~~~~~~~~~~~~~ \\
~~${T_\text{syst}}$ &  &10.2\%~~~~~~~~~~~~~~~~~ \\
\hline 
~~Total          &  &11.0\%~~~~~~~~~~~~~~~~~ \\  
\hline 
\hline
\end{tabular}\label{tab:SysUncert} 
\end{table}
       
\par In summary, an extended search for axion dark matter is performed using the CAPP18T axion haloscope. The detector was fully reconstructed after the 18\,T magnet quenching incident. The thermal contact of the detector components and the impedance configuration of the {rf} chains were improved during the rebuilding of the detector. Compared to {run 1}, the ${T_\text{syst}}$ of {run 2} is lowered by $\sim$35\,\%. The $\beta$ is maintained at $\sim$2. The new results set the strongest limits for excluding the KSVZ benchmark model axion-photon-photon couplings in the axion mass range of 19.883 -- 19.926\,$\mu$eV (90\,\% C.L.). The results demonstrated that a reliable axion dark matter search is possible using the innovative techniques of the {high-temperature superconducting} magnet and the JPC quantum amplifier.\\

\begin{acknowledgements}
\par This research is supported by the Institute for Basic Science (IBS-R017-D1-2021-a00/IBS-R017-G1-2021-a00). This work is also supported by the New Faculty Startup Fund from Seoul National University. We also thank the team of the Superconducting Radio Frequency Testing Facility at Rare Isotope Science Project at IBS for sharing the experiment space. 
\end{acknowledgements}
\bibliography{capp18t}
\end{document}